\documentclass[a4paper,11pt]{article}
\usepackage{times}
\usepackage{amsmath,amssymb,amsfonts,amsthm}
\def\be{\begin{equation}}
\def\ee{\end{equation}}

\newcommand{\sfrac}[2]{{\textstyle{#1\over#2}}}

\begin{document}
\title{\sc 
The Trace-Free Einstein Equations and inflation}
\author{
{\sc George F. R. Ellis}${}^{1}$\thanks{E--mail: {\tt
George.Ellis@uct.ac.za}}.\\ \\
Mathematics Department and ACGC, University of Cape Town; \\
Trinity College and DAMTP, Cambridge}

\maketitle
\begin{abstract}
The trace-free version of the Einstein Gravitational equations,
essentially equivalent to unimodular gravity, can solve the
troubling issue of the huge discrepancy between quantum field theory
estimates of the vacuum energy density and the astronomically
observed value of the cosmological constant. However it has been
suggested that this proposal cannot work because it prevents the
inflaton potential energy from driving inflation. It is shown here
that that concern is unjustified: inflation proceeds as usual if we
adopt the trace free gravitational equations.
\end{abstract}
\section{Introduction}
The motivation for this brief paper was a statement by M R Douglas
at a recent
meeting\footnote{http://www.damtp.cam.ac.uk/events/infinities2013/}
that a reason for proposing string theory/M-theory as a theory of
gravity is the otherwise intractable problem of vacuum energy: a
major discrepancy between quantum field theory and Einstein's
gravitational theory. However it is known \cite{Wei89,Elletal11}
that  there is a much simpler alternative explanation for this
problem: it can be solved by using the Trace-Free Einstein Equations
(`TFE') instead of the standard Einstein field equations (`EFE').
The issue leading to doubt about this proposal is that it is not
obvious it will work in the case of inflationary cosmology (M
Douglas, private communication). The aim of this paper is to clarify
that issue. There are of course other reasons for proposing string
theory (in particular, it promises to be a fundamental theory
unifying the forces of nature)
which are not affected by this argument; but the issue raised by Douglas can indeed be resolved.\\

General relativity leads us to expect  that the gravitational effect
of a quantum vacuum will be the equivalent of an effective
cosmological constant~\cite{zeldo}, which will cause an accelerated
expansion of the universe through the Raychaudhuri equation
\cite{EllMaaMac12}. However, estimates based in Quantum Field Theory
(QFT) of the magnitude of the vacuum energy exceed by at least 60
orders of magnitude the value of the cosmological constant as
estimated from astronomical observations of distant
supernovae~\cite{Wei89,Car01}. This blatant contradiction with
observations indicates a profound discrepancy between  General
Relativity (GR) and Quantum Field Theory. This  is a major problem
for theoretical physics, because these are two of our most
successful physical theories, and the problem does not occur in a
regime where we would expect either theory to break down. We need to
resolve that discrepancy, no matter what final quantum gravity
theory is adopted. Carroll and Mersini \cite{CarMer01} state the
options as follows:

\begin{quote}
\emph{The fact that the observed cosmological constant is much smaller than the expected value
may provide a crucial clue in our attempts to understand the nature of spacetime.
Most attempts to solve the problem can be characterized as making the vacuum energy
much smaller than its natural value. Alternatively, however, we can imagine keeping a large
vacuum energy, but changing the gravitational dynamics in such a way that the vacuum
does not act as a (significant) source of spacetime curvature. Since it is only through its
gravitational influence that the vacuum energy can be measured, such an arrangement could
reconcile the naive estimates $\rho_{vac} \geq (10^{18} GeV)^4$ with the observationally favored result
$\rho_{vac} \sim (10.3 eV)^4.$}
\end{quote}
The first route to resolution can be attempted by claiming cancelations between the various
contributions to vacuum energy will occur, but this is very implausible, because the energies of
the different fields that contribute (see estimates  (\ref{vac1})-(\ref{vac3}) below) would have
to cancel to extraordinarily high accuracy (to over 100 decimal places). This is not a plausible
explanation. One proposed solution is the conjunction of the idea of the landscape of vacua of M-theory
with the idea of a multiverse \cite{Sus03}; hence this can be proposed as a motivation for the correctness of string theory, as noted by Douglas.\\

The second route for solving the problem can be attained by a method summarized by Weinberg in his review \cite{Wei89}. One uses a Trace-Free version of the Einstein Equations instead of the standard General Relativity field equations. If these trace-free equations are adopted, which is effectively the same as adopting unimodular gravity \cite{AndFin71,FinGalBau00,Unr89,Smo09}, the vacuum energy has no gravitational effect \cite{Elletal11}. This does not determine a unique value for the effective cosmological constant, but it does solve the huge discrepancy between theory and observation in the standard approach; thus it resolves an otherwise disastrous discrepancy between two key areas of theoretical physics. \\

However, there is a problem for cosmology. Carroll and Mersini
\cite{CarMer01} state it thus:
\begin{quote}
\emph{An obvious question which arises in any model which would make the metric insensitive
to vacuum energy is, how are we to explain the apparent nonzero value of the cosmological
constant today, or implement an inflationary scenario in the early universe?
}\end{quote}
There are two paradoxes associated with the use of this approach in the cosmological context. Firstly the inflaton potential $V(\varphi)$ does not enter the trace free cosmological equations, so it is unclear how it can affect cosmological expansion, as is assumed in the standard theory. The resolution is explained in Section \ref{sec:paradox1}: although it is not explicitly used in any of the trace free Einstein equations, it enters the dynamics via the energy conservation equation. But this leads to a second paradox; if it is the case that the scalar potential re-enters the effective Einstein equations via this indirect route, why does the same not also apply to the vacuum energy? The resolution is explained in Section \ref{sec:paradox2}. This still leaves a puzzle: the absolute value of the potential does not occur in the dynamical equations of cosmology, but it does determine important aspects of inflationary dynamics. How this happens is explained in Section \ref{sec:absolute}, and is related to the present day value of the cosmological constant in Section \ref{sec:remnant}. \\

This paper builds on a previous paper \cite{Elletal11},\footnote{Sections \ref{sec:basic} and \ref{sec:tfe} here summarise part of that paper.} which describes this solution to the vacuum energy issue in general terms, but does not explicitly deal with these paradoxes. I do not deal with the issue of obtaining a variational principle for unimodular gravity (what Lagrangian gives it): that is dealt with in various other papers, see  \cite{AndFin71,Unr89,AlvFae07,Smo09,Alv12}. Other ways of justifying a unimodular theory are proposed in scenarios given in \cite{Pad08,KimPad12}. Conformally invariant theories \cite{Mal11,LuPanPop11} can also be implemented in a unimodular way, as the metric determinant can be set to unity by using the conformal transformation. Variational principles for a cosmologically viable class of such theories are given in \cite{KalLin13}.\\

\section{The Basic Problem}\label{sec:basic}
\subsection{Gravitational dynamics}\label{sec:dyn}
Classical gravitational dynamics is governed by the Einstein Field Equations
(EFE)
\begin{equation}
\label{EFE}
R_{ab} - \sfrac{1}{2}\,Rg_{ab}
+ \Lambda g_{ab}= 8\pi G\,T_{ab} \ ,
\end{equation}
where $R_{ab}$ is the Ricci tensor, $R$ the Ricci scalar, $g_{ab}$ the metric tensor, $T_{ab}$ the matter energy-stress tensor, $G$ the gravitational constant, and  $\Lambda$ the cosmological constant. These are subject to the consistency relations
\begin{equation}
\label{EFE_div}
\nabla^{b}(R_{ab} - \sfrac{1}{2}\,Rg_{ab}) = 0
\end{equation}
which guarantee consistency of the time development of the EFE.  Applied to (\ref{EFE}), and assuming $\Lambda$ is indeed a constant,
they imply the conservation equations
\begin{equation}\label{cons_em}
\nabla^{b}T_{ab}=0 \ ,
\end{equation}
for the total energy--momentum tensor.\\

In standard cosmology \cite{PetUza09,EllMaaMac12}, the metric tensor is
assumed to take the spatially homogeneous and isotropic Robertson-Walker form
\begin{equation}
\label{RW metric} ds^2 = -dt^2 + a^2(t)d\sigma^2 \ ,
\qquad
u^a = \delta_{0}{}^{a} \ ,
\end{equation}
with $a(t)$ a time-dependent scale factor, $d\sigma^2$ the
metric of a 3-space of constant curvature $k$ ($k = +1,\, 0,$ or $-1$), and $u^{a}$ the
normalised matter 4-velocity field ($u_{a}u^{a} = -1$). The Hubble parameter is $H(t) := \dot{a}/a$. The energy--momentum tensor of
the matter takes a perfect fluid form:
\begin{equation}
\label{PF}
T_{ab} = (\rho + p)u_a u_b + p\,g_{ab} \ ,
\end{equation}
where the matter mass density $\rho$ and the matter isotropic
pressure $p$  are related through an equation of state
$p =p(\rho)$
(which may be defined parametrically, e.g. $\rho=\rho(\varphi,d\varphi/dt)$, $p=p(\varphi,d\varphi/dt)$ as in (\ref{eq:rho}) below). The conservation equations (\ref{cons_em}) reduce to the
mass--energy conservation relation
\begin{equation}
\label{cons1}
\dot{\rho} + 3\,\frac{\dot{a}}{a}\left(\rho+p\right)=0 \ ,
\end{equation}
where a dot denotes derivation with respect to proper
time $t$. For the metric (\ref{RW
metric}), the EFE (\ref{EFE}) reduce to two non-trivial equations:
the Raychaudhuri equation
\begin{equation}
\label{ray}
3\,\frac{\ddot{a}}{a} + 4\pi G\left(\rho+3\,p\right)
- \Lambda  = 0 \ ,
\end{equation}
and the Friedman equation
\begin{equation}
\label{fried}
3\left(\frac{\dot{a}}{a}\right)^{2} = 8\pi G\rho
+\Lambda  -\,\frac{3k}{a^2}  .
\end{equation}
Because (\ref{cons1}) implies
$ d(\rho a^2)/dt = - a \dot{a} (\rho +3p),$
 on multiplying (\ref{ray}) by an integrating factor $a \dot{a}$, equation (\ref{fried}) is a first integral of (\ref{cons1}) and (\ref{ray}); indeed this has to be the case, or the dynamics would be over-determined. Hence we have the result:

\begin{quote}
\textbf{Two effective cosmological equations}: \emph{Given the equation of state,
any two of (\ref{cons1})- (\ref{fried}) imply the third. When they are satisfied, all 10 EFE (\ref{EFE}) are satisfied.}
\end{quote}
\subsection{Vacuum energy}

From the quantum field theory side, vacuum energy \cite{Car01} has an effective pressure $p_{vac}$ and energy density $\rho_{vac}$ that give an effective cosmological constant $\Lambda$. On  moving the $\Lambda$-term to the right hand side of (\ref{EFE}), it is equivalent to a perfect fluid (\ref{PF}) with equation of state
 $ \rho_{\Lambda} = - p_{\Lambda} = \frac{\Lambda}{8\pi G}$\,,
implying
 $ \rho_{\Lambda} + 3 p_{\Lambda} = -  \frac{\Lambda}{4\pi G}$
(which also follows from (\ref{ray})).
This will cause acceleration rather than deceleration if $\Lambda > 0$. However in what follows, the contribution of the vacuum energy to the field equation (\ref{EFE})  will not be regarded as a fluid, but as an effective cosmological constant term
  $\Lambda_{vac} g_{ab}$,
which is Lorentz invariant because the vacuum does not pick out any specific timelike direction.\\

Amongst the contributions to vacuum energy \cite{Car01} are that from the electroweak field
\begin{equation}\label{vac1}
\rho^{(EW)}_{\textsf{vac}} \simeq (200\, GeV)^4 \simeq 3 \times 10^{47} erg/cm^3,
\end{equation}
that from quantum chromodynamics
\begin{equation}\label{vac2}
\rho^{(QCD)}_{\textsf{vac}} \simeq (0.3\, GeV)^4 \simeq 1.6 \times 10^{36} erg/cm^3,
\end{equation}
and that from fluctuations at the Planck scale
\begin{equation}\label{vac3}
\rho^{(Pl)}_{\textsf{vac}} = M_{pl}^4 \simeq (10^{19}\, GeV)^4 \simeq 2 \times 10^{110} erg/cm^3.
\end{equation}
These will all contribute to the total effective cosmological constant term in (\ref{EFE}):
\begin{equation}\label{cctot}
  \Lambda_{\textsf{eff}}\, g_{ab} = (\Lambda +  \Lambda_{\textsf{vac}})\, g_{ab} =  (\Lambda + 8 \pi G \sum_{(i)} \rho_{\textsf{vac}}^{(i)}) \,\, g_{ab}
\end{equation}
where the sum is over all such vacuum energy contributions. The total term is still Lorentz invariant.

\subsection{The Problem}
The cosmologically observed value, determined through supernovae observations in conjunction with studies of large scale structure formation and analysis of the cosmic microwave background anisotropy spectrum \cite{Car01,PetUza09,EllMaaMac12},  is
\begin{equation}\label{lew}
0 < \rho^{(obs)}_\Lambda \leq (10^{-12}\, GeV)^4 \simeq 2 \times 10^{-10} erg/cm^3,
\end{equation}
in strong contradiction to the estimates (\ref{vac1})-(\ref{vac3}). This clearly indicates a severe problem. Cancellations between terms such as (\ref{vac1}) and (\ref{vac2}) to give (\ref{lew}) is clearly extraordinarily unlikely (a symmetry principle such as supersymmetry might provide term by term cancellation; but even if supersymmetry applies to the underlying fundamental physics, it is a broken symmetry that does not hold in the effective theories that hold sway in the observed universe).

\begin{quote}
\textbf{The vacuum energy disaster:} \emph{The Raychaudhuri equation (\ref{ray}) shows that $(\rho +
3p$) is the active gravitational mass density, while the Friedman Equation (\ref{fried}) shows the energy equation for expansion has $\rho$ as source term. If we include the vacuum energy densities (\ref{vac1}) - (\ref{vac3}) as source terms in the Einstein equations (\ref{EFE}), both equations (\ref{ray}) and (\ref{fried}) are a disaster}.
\end{quote}

\section{Trace-free Einstein gravity}\label{sec:tfe}
A proposal that solves the basic problem is to take the trace-free part of the EFE to
get the Trace-Free Einstein Equations (TFE). This is a subset of
the EFE that, on using the energy conservation equations (\ref{cons_em}), give back the full standard EFE (\ref{EFE}) but with the cosmological constant in that equation an integration constant that has nothing to do with vacuum energy (see \cite[pp.~11--13]{Wei89} and references given there.)
More recently, it has been developed under the name of ``unimodular gravity'', see
\cite{AndFin71,Unr89,FinGalBau00,Smo09} and references therein.

\subsection{The Trace-Free Equations}\label{sec:TFE}
We use a hat to denote the trace-free part of a symmetric tensor:
so
$$\hat{G}_{ab} := R_{ab} -\sfrac{1}{4}\,R g_{ab} \ ,
\quad
\hat{T}_{ab} := T_{ab} -\sfrac{1}{4}\,T g_{ab}
\quad\Rightarrow\quad
\hat{G}_{a}{}^a = 0 \ ,
\quad
\hat{T}_{a}{}^a = 0 \ .$$
On taking its trace-free part, the EFE
(\ref{EFE}) implies the TFE
\begin{equation}
\label{EFE_TF1}
R_{ab} -\sfrac{1}{4}\,R g_{ab} =
8\pi G \left(T_{ab} -\sfrac{1}{4}\,T g_{ab}\right) \quad\Leftrightarrow\quad
\hat{G}_{ab} = 8\pi G\,\hat{T}_{ab}.
\end{equation}
It is crucial that both sides of the equation have the same symmetry.
We adopt these as the gravitational field equations, instead of
(\ref{EFE}).
The conservation relation (\ref{cons_em}) for the
energy--momentum tensor $T_{ab}$
is then no longer a consequence of the geometrical identity (\ref{EFE_div}): it is to be introduced as a
{\em separate assumption\/}.\\

 Thus the starting point is assuming that both (\ref{cons_em}) and (\ref{EFE_TF1}) hold. There are 9 field equations; the energy--momentum tensor of the matter can have a non-zero trace, but only the trace-free
part gravitates. As discussed in \cite{Wei89,Elletal11}, the theory is subject to an {\em integrability condition\/}:
differentiating (\ref{EFE_TF1}), and using (\ref{EFE_div}) and
(\ref{cons_em}) gives
\begin{equation}
\label{EFE_TF2}
\nabla_{b}\left(R^{ab} -\sfrac{1}{4}\,R g^{ab}\right)
= 8\pi G\,\nabla_{b}\left(T^{ab}
-\sfrac{1}{4}\,T g^{ab}\right)
\quad\Rightarrow\quad
\nabla_{a}R = -\,8\pi G\,\nabla_{a}T \ .
\end{equation}
Integrating, $\displaystyle\left(R+8\pi G\,T\right)$ is a constant:
 $\hat{\Lambda} := (R+ 8\pi G\,T)/4 \, \Rightarrow \, \nabla_{a}\hat{\Lambda} =0.$
Substituting into (\ref{EFE_TF1}) to eliminate $T$ gives
back (\ref{EFE}), but with a
new effective cosmological constant:
\begin{equation}
\label{EFE3}
R_{ab} -\sfrac{1}{2}\,R g_{ab} + \hat{\Lambda}g_{ab}
= 8\pi G\,T_{ab} \ .
\end{equation}
So the way it works is as follows:
\begin{quote}
\textbf{The Trace Free Equations}: \emph{instead of using the EFE (\ref{EFE}), we assume both the TFE
(\ref{EFE_TF1}) and the matter conservation equations
(\ref{cons_em}) hold. The integrability condition (\ref{EFE_TF2})
follows from these equations. Integrating gives (\ref{EFE3})}.
\end{quote}
\subsection{The vacuum energy decoupled from spacetime curvature}
We have a remarkable result \cite{Wei89}:
\begin{quote}
\emph{
\textbf{Disempowering the vacuum energy}: the TFE (\ref{EFE_TF1}) together with the differential relations
(\ref{cons_em}) are functionally equivalent to the EFE (\ref{EFE3}), but with an effective cosmological constant  $\hat{\Lambda}$ that is unrelated to the vacuum energy $\Lambda_{\rm vac}$.}
\end{quote}
The gravity theory based on the TFE recovers all the vacuum solutions of the EFE unchanged, so e.g. results from the Schwarzschild and Kerr black hole solutions are still valid. However, cosmology no longer has a vacuum energy problem, as  the energy densities (\ref{vac1})-(\ref{vac3}) do not occur in (\ref{EFE_TF1}); hence they cannot affect spacetime curvature.\\

As a specific example, for a perfect fluid (\ref{PF}), the matter
source term is the manifestly trace-free energy--momentum
tensor
$\hat{T}_{ab} = \left(\rho+p\right)
\left(u_a u_b + \sfrac{1}{4}\,g_{ab}\right) \, \Rightarrow \,\,\hat{T}^a_{\,\,\,\,a} = 0.
$ 
The fluid density and pressure enter the field equations only in terms of the
inertial mass density $(\rho+p)$, which vanishes for a vacuum. The vacuum energy does not affect spacetime curvature.

\section{Scalar fields in a FLRW spacetime}\label{sec:sfcosm}
Inflationary cosmology is driven by one or more effective scalar fields $\varphi(x^i)$.
The dynamics of the field is governed by the Klein Gordon Equation
\begin{equation}\label{KGequiv}
\ddot\varphi+3H\dot\varphi
+\sfrac{dV}{d\varphi} = 0 .
\end{equation}

The stress energy tensor of a scalar field $\varphi$ with potential $V(\varphi)$ is
\begin{equation}
T_{ab}=\partial_{a}\varphi\partial_{b}\varphi - \sfrac{1}{2}\left[\,\partial_{c}\varphi\partial^{c}\varphi
+2V(\varphi)\,\right]g_{ab}
\end{equation}
For the FLRW case (\ref{RW metric}) with scale factor $a(t)$ and Hubble parameter $H(t) = \dot{a}(t)/a(t)$, the spacetime symmetries imply the scalar field will depend only on time: $\varphi = \varphi(t)$. It's gradient $\partial_{a}\varphi$ will relate to the fundamental 4-velocity $u^a$ by
$u_a = -\sfrac{1}{\dot{\varphi}}\partial_{a}\varphi \Leftrightarrow \partial_{a}\varphi = - \dot{\varphi} u_a, $
and so  its stress-tensor (\ref{phistress}) will take the perfect fluid form
\begin{equation}\label{phistress}
T_{ab}=  (\dot{\varphi})^2 u_a u_b +\left[\sfrac{1}{2}(\dot{\varphi})^2)
- V(\varphi)\,\right]g_{ab}\,.
\end{equation}
The effective energy density $\rho$ and the pressure $p$ are
\begin{equation}\label{eq:rho}
\rho = T_{ab}u^au^b = \sfrac{1}{2}(\dot{\varphi})^2 + V(\varphi)\,,\,
 p = \sfrac{1}{3}T_{ab} h^{ab}=\frac{1}{2}(\dot{\varphi})^2 -V(\varphi)\,.
\end{equation}
This gives an active gravitational mass density $\rho_{grav}$ and inertial mass density $\rho_{inert}$ as follows: \begin{equation}
\rho_{\textsf{grav}} := \rho +3p = 2\dot{\varphi}^{2}-2V(\varphi),\,\,  \label{scalar1}
\rho_{\textsf{inert}} : =(\rho +p) = (\dot{\varphi})^2\,.
 \end{equation}

The mass--energy conservation relation is (\ref{cons1}).
Calculating $\dot{\rho}$ in two ways,
\begin{equation}
  \dot{\varphi} \left(\ddot{\varphi} +\sfrac{d V(\varphi)}{d \varphi}\right)
\,\, \underbrace{=}_{(\ref{eq:rho})}\,\, \dot{\rho} \underbrace{=}_{(\ref{cons1})} - 3\,H (\rho +p)\,\, \underbrace{=}_{(\ref{scalar1})}\,\, - 3\,H(\dot{\varphi})^2
\end{equation}
giving
\begin{equation}\label{refff}
\dot{\varphi} \left(\ddot{\varphi} +\sfrac{d V(\varphi)}{d \varphi} + 3\,H \dot{\varphi}\right) = 0
\end{equation}
So if $\dot\varphi \neq 0$,  
energy conservation yields the
Klein Gordon equation (\ref{KGequiv}), which is the dynamic equation for the scalar field.\\

\begin{quote}
\textbf{Generic case:} \emph{If $\dot\varphi \neq 0$, the Klein Gordon equation (\ref{KGequiv}) and conservation equation (\ref{cons1}) are equivalent. Either one implies the other.
}\end{quote}
\begin{quote}
\textbf{Exceptional case:} \emph{If $\dot\varphi = 0$, the Klein Gordon equation (\ref{KGequiv}) shows that $\sfrac{dV}{d\varphi} = 0$: the potential is a constant, that is, we have a cosmological constant rather than a dynamic field. 
Equations (\ref{eq:rho})  show $\rho = -p  = V(\varphi) = \textrm{const}$, and (\ref{cons1}) is satisfied identically.}\end{quote}

\section{TFE cosmology for a scalar field:}
When the TFE equations are used in inflationary cosmology, one assumes the metric (\ref{RW metric})
and the trace free field equations (\ref{EFE_TF1}) with matter source term that due to a scalar field
(\ref{phistress}) with its dynamics subject to the Klein Gordon equation (\ref{KGequiv}). The question
is, can have slow roll inflation in the case of unimodular gravity?

\subsection{The first paradox and its resolution}\label{sec:paradox1}
The trace-free part of the scalar field stress tensor (\ref{phistress}) is
$\hat{T}_{ab}=\partial_{a}\varphi\partial_{b}\varphi
-\sfrac{1}{4}(\partial_{c}\varphi\partial^{c}\varphi)g_{ab}\, ,$
whatever the potential. This is the source of the TFE equations (\ref{EFE_TF1}) that govern the dynamics of cosmology. Following the prescription above (Section \ref{sec:TFE}), the equations governing the dynamics of cosmology are the conservation equation
\begin{equation} \label{cons11}
 \dot{\rho} = - 3\,\frac{\dot{a}}{a}\left(\rho+ p\right) \,= - 3 H \, (\dot{\varphi})^2\, ,
\end{equation}
and the trace-free field equation
\begin{equation}
\label{FL_TF}
\frac{\ddot{a}}{a} - \left(\frac{\dot{a}}{a}\right)^{2}
-\frac{k}{a^{2}}
= -\,4\pi G\left(\rho+p\right) = -\,4\pi G (\dot{\varphi})^2\ .
\end{equation}
Although it has some similarities, this is not the same as the
dynamics of the self-tuning equations (6), (88) of Carroll and
Mersini \cite{CarMer01}, even though there are similarities in that
their equations also depend only on $(\rho +p).$ Their equation (88)
is
equivalent to 
$\frac{\ddot{a}}{a}  = \frac{2\pi G}{9}(\rho +p)$, in contrast to the above.\\

The matter terms in this case only affects space-time curvature via the inertial mass density $\rho_{\textsf{inert}}$, so the gravitational mass density $\rho_{\textsf{grav}}$ - which enters the key Raychaudhuri equation governing the gravitational dynamics \cite{Ell71} --- does not explicitly enter the cosmological equation (\ref{FL_TF}). Neither does the inflaton potential energy $V(\varphi)$.
\begin{quote}
\textbf{The first paradox}: \emph{The potential term $V(\varphi)$ does not occur in the dynamical equation (\ref{FL_TF}) for cosmology, and so it does not directly affect the spacetime curvature, when the TFE equations (\ref{EFE_TF1}) are adopted as the gravitational equations. It therefore apparently has no gravitational effect.}
\end{quote}
So how does the scalar field potential energy affect the outcome seeing it is not present in the gravitational equation (\ref{FL_TF})?

\begin{quote}
\textbf{The resolution}\emph{ is that the potential term occurs implicitly through the time derivative on the left of the conservation equation (\ref{cons11}).}
\end{quote}
By (\ref{eq:rho}), that equation is  actually
\begin{equation}
\label{FL_TF5}
\partial\left(\sfrac{1}{2}(\dot{\varphi})^2 + V(\varphi)\,\right)/\partial t = - 3 H \, (\dot{\varphi})^2\, ,\end{equation}
 so the time derivative of  $V(\varphi)$ enters through this equation.
This is where the potential comes in.
This leads to the standard inflationary dynamics, because (section \ref{sec:dyn}) any two of the three equations (\ref{ray}), (\ref{cons1}), (\ref{fried}) imply the third. Thus the  two equations (\ref{cons11}), (\ref{FL_TF}), with (\ref{FL_TF}) obtained as a linear combination of (\ref{ray}) and (\ref{fried}), are equivalent to the standard dynamics of inflationary cosmology, but with an effective cosmological constant $\hat{\Lambda}$.
To see this, multiply (\ref{FL_TF}) by an integrating factor $(6\dot{a}/a)$, use the Klein Gordon equation (\ref{KGequiv}), and integrate to get the Friedman equation
\begin{equation}
\label{friedsf}
3\left(\frac{\dot{a}}{a}\right)^{2} +\,\frac{3k}{a^2} -\hat{\Lambda}   = 8\pi G  \left( \sfrac{1}{2}(\dot{\varphi})^2 +V(\varphi)\right) 
\end{equation}
where $\hat{\Lambda}$ is a constant of integration.
Substituting back into (\ref{FL_TF}) will give the  effective Raychaudhuri equation
 \begin{equation}
\label{raysf}
3\,\frac{\ddot{a}}{a}  - \hat{\Lambda}   =  -4\pi G \left(2\dot{\varphi}^{2}- 2V(\varphi)\right).
\end{equation}
 Now $V(\varphi)$ does indeed appear in both (\ref{friedsf}) and  (\ref{raysf}) and does affect spacetime curvature.
This result can alternatively be checked by taking the time derivative of (\ref{friedsf}) and using (\ref{KGequiv}) to get (\ref{FL_TF}), whatever the value of $\hat{\Lambda}$.

\subsection{The second paradox and its resolution}\label{sec:paradox2}
But now the second paradox arises:
\begin{quote}
\textbf{Second paradox}: \emph{If this works for $V(\varphi)$, why does the same not happen for the vacuum energy $\Lambda_{vac}$? Why does this huge energy not re-appear, Phoenix like, in the effective EFE, as happens for $V(\varphi)$?}
\end{quote}
The point here is that the causal chain whereby $V(\varphi)$ enters the equation only works when $d\varphi/dt \neq 0$.
When $d\varphi/dt =0$,  the term  $V(\varphi)$ is constant and drops out of (\ref{FL_TF5}). That is,
\begin{quote}
\textbf{Non-reappearance of vacuum energy $\Lambda_{vac}$}  \emph{in the effective field equations is because it is a constant, and so has no effect in (\ref{FL_TF5}). This is the exceptional case of Section \ref{sec:sfcosm}, corresponding to $\dot{\varphi} = 0$, where the conservation equations are not equivalent to the Klein Gordon equation }.
\end{quote}

\subsection{Puzzle: The zero point of the potential $V(\varphi)$}\label{sec:absolute}

However there is still a significant puzzle to resolve: as just shown, the dynamic equations (\ref{cons11}), (\ref{FL_TF}) for inflationary cosmology in the TFE approach do not involve $V$, and the equations (\ref{FL_TF5}), (\ref{KGequiv}) that lead to its appearance only involve its derivative and so are insensitive to the normalisation of the potential $V(\varphi)$: these equations are invariant under the change
\begin{equation}\label{vchange}
  V(\varphi) \,\mapsto \,V(\varphi) + C
\end{equation}
for arbitrary constants $C$. However $V(\varphi)$ does directly affect the gravitational
dynamics given by (\ref{friedsf}), (\ref{raysf}).  In particular, in the slow roll regime of
inflation where $\varphi = \varphi_{\textsf{sr}} \simeq const$, we can ignore
$\dot{\varphi}^{2}$ and the curvature term $k/a^2$,  and  on defining
\begin{equation}\label{slowroll}
V_{\textsf{sr}} :=   V(\varphi_{\textsf{sr}}) + \hat{\Lambda}/(8\pi
G )
\end{equation}
we get the slow roll inflation equations
\begin{equation}
\label{rayinf}
3\,\frac{\ddot{a}}{a}    
= 8\pi G V_{\textsf{sr}}\,,\,\,
3\left(\frac{\dot{a}}{a}\right)^{2}  
= 8\pi G V_{\textsf{sr}}.
\end{equation}
The slow roll parameters for the inflaton field $\varphi$ are determined by $V'/V$ and $V''/V$, which are invariant under (\ref{vchange}). However the  inflationary power spectra of tensors and scalar field fluctuations depend on $H^2 := (\dot{a}/a)^2$ which depends directly on $V_{\textsf{sr}}$, as is shown in the second of (\ref{rayinf}):
 $P_{GW},\, P_{\delta \varphi}\, \propto H^2_*/k^3$
where * denotes Hubble crossing, so these spectra are determined by $V$ (at Hubble crossing). So the actual  value of $V(\varphi)$ does indeed appear in the effective equations, and makes a difference to the cosmological outcomes such as the tensor to scalar ratio. How does this happen if only derivatives of $V(\varphi)$ occur in the link (\ref{FL_TF5}) to cosmological dynamics mentioned in Section \ref{sec:paradox1}?
\begin{quote}
  \textbf{Puzzle}: \emph{While both (\ref{FL_TF5}) and (\ref{KGequiv}) do depend on $V(\varphi)$, which therefore determines the motion, they contain only the derivative $dV
  /d\varphi$, not the value  $V(\varphi)$. But this value does indeed affect cosmological dynamics, as for example is shown in (\ref{rayinf}).}
\end{quote}
To make this explicit, we can consider the case of a quadratic potential
 $ V(\varphi) = V_0 + \frac{1}{2} M^2 \varphi^2$,
where $V_0$ and $M^2$ are constants. Then $V_{\textsf{sr}}$ in (\ref{rayinf}) is given by $V_{\textsf{sr}} = V_0 + \frac{1}{2} M^2 \varphi_{\textsf{sr}}^2+ \hat{\Lambda}/(8\pi G )$, and (\ref{friedsf})
 indeed is affected by $V_0$.

The resolution is that there are \emph{two} constants 
in this equation, namely $\hat{\Lambda}$ (a constant arising from integrating the gravitational field equations) and $V_0$; 
but they are dynamically indistinguishable from each other, as their effects on the cosmological evolution are identical. We can incorporate either one into the other to give the one essential free constant in the dynamics, for example we can define
\begin{equation}\label{v0}
\hat{\Lambda}_{\textsf{eff}} = \hat{\Lambda} + 8\pi G V_0
\end{equation}
so (\ref{friedsf}) becomes
\begin{equation}
\label{frieds3}
3\left(\frac{\dot{a}}{a}\right)^{2} +\,\frac{3k}{a^2} =\hat{\Lambda}_{\textsf{eff}}  + 8\pi G  \left( \sfrac{1}{2}(\dot{\varphi})^2 + \frac{1}{2} M^2 \varphi^2 \right).
\end{equation}
Just as is the case in the rest of theoretical physics, it is the variation of potential energy that matters, not its specific value. This can be set arbitrarily without determining the dynamics,
because given any choice $V_0$, 
the integration constant $\hat{\Lambda}$  determines the physical outcome.

\begin{quote}
\textbf{Resolution}: \emph{In the Friedman eqn (\ref{friedsf}), the
potential occurs in the energy density of the scalar field $\rho =V
+ \dot{\varphi}^2/2$. However the zero point potential $V_0$ is
indistinguishable from the integration constant $\hat{\Lambda}$ and
the arbitrariness of the one can be traded off in terms of the
arbitrariness of the other. What actually determines the dynamics is
$\hat{\Lambda}_{\textsf{eff}}$ given by (\ref{v0}).}
\end{quote}
Consequently, the dynamics during inflation of a universe governed
by the TFE will be the same as in standard GR, except that the zero
point of the potential does not control the dynamics. You have the
freedom to vary an integration constant in the dynamical equations,
which is equivalent to a shift in the value of the minimum of the
potential.

\subsection{The issue of remnant energy}\label{sec:remnant}
A final question is, how does this all relate to the value of the cosmological constant determined today? \\

At the end of inflation, the scalar field will die away and get
converted to matter and radiation. The field will settle to its
state of lowest energy: $\varphi \rightarrow 0$ and
$V(\varphi)\rightarrow V(0)$, which remains today (the constant
offset $V_0$ is not affected by reheating, i.e. does not decay).
However it is not the only effective term: $\hat{\Lambda}$ also
remains, and they both contribute as before.
\begin{quote}
\textbf{The present value}: \emph{the cosmological constant that remains today is just
$\hat{\Lambda}_{\textsf{eff}}$ given by (\ref{v0}). That will be the value measured today}.
\end{quote}
From the viewpoint of the physics, it is an arbitrary integration
constant that can be fitted to the observed value of the
cosmological constant as determined by the concordance model of
cosmology \cite{PetUza09,EllMaaMac12}.\\

So how does this value relate to the slow roll value during
inflaton? The slow roll value then was
\begin{eqnarray}\label{today}
V_{sr} &=&   \left( V(\varphi_{\textsf{sr}})-V_0\right) \,+\,
\left(V_0+ \frac{\hat{\Lambda}}{8\pi G }\right)\nonumber\\
&=& \Delta V \,+
\, \frac{\hat{\Lambda}_{eff}}{8\pi G }
\end{eqnarray}
where $\Delta V:=(V(\varphi_{\textsf{sr}})-V_0)$ is the difference
between today's value of $V(\varphi)$ and the value it had during
inflation. This term dominated then, but dies away at the end of
inflation, leaving behind the effective cosmological constant
$\hat{\Lambda}_{eff}$ that we observe today. It is $\Delta V$ that
drove the inflationary expansion, not $V_0$. The end of inflation
was marked by the transition
\begin{equation}\label{end}
 \Delta V \rightarrow 0\,\,\Rightarrow\,\,V \rightarrow \frac{\hat{\Lambda}_{eff}}{8\pi G
 }
\end{equation}
 and $\hat{\Lambda}_{eff}$ has been the effective cosmological constant from then on to
 the present day. If $V_0$ is a false minimum rather than the real minimum, a
further such transition may take place in the future.\\

The point is that there is
\begin{itemize}
  \item one integration constant from the field equations (going from the TFE to the EFE), plus
  \item one further integration constant for each scalar field that may be dynamically important (going from the conservation equation to the Klein Gordon equation for that field).
\end{itemize}
It is the sum of these terms that gives the present day effective
cosmological constant. The inflationary value of the potential was
$\Delta V$ larger than today, just as in standard inflationary
theory.

\subsection{Issues that arise} The following questions have been raised as regards this scenario:
\begin{itemize}
  \item The limit from slow roll, $\epsilon >0$, to a cosmological constant, $\epsilon =0$ seems not to be
  continuous.

  - The dynamics is not discontinuous as $\epsilon \rightarrow 0$: rolling on a flat potential will go smoothly
  to rolling on a sloping potential. What is discontinuous is the
  dependence relation of the effective constant to the potential absolute value. There will be no
  problem in this regard provided the potential is not flat
  \emph{everywhere}; the method above will give the desired link to
  dynamics provided there is \emph{some} domain where $dV/d\varphi \neq 0$.
  \item It is unsatisfactory that the Hubble parameter, a physical quantity, depends on the integration
    constant $\hat{\Lambda}$.

    - That is the case in standard cosmology, where at late times
    the expansion rate depends on $\Lambda$. One does not have to
    think of it as an integration constant: it can be taken as just
    one more constant of nature in the combined standard model
    of cosmology and particle physics.
  \item Is this compatible with nucleosynthesis?

  - Yes, the value of the cosmological constant then was $\hat{\Lambda}_{eff}$ just as
 it is today. The standard compatibility of that value with
 cosmological observations applies.
\end{itemize}
\section{Viability of the TFE equations}
If the true gravitational field equations are the TFE  (\ref{EFE_TF1}), implying that
only the trace-free part of the energy--momentum tensor $T_{ab}$ of
matter is gravitating, then the effective cosmological constant
$\hat{\Lambda}_{\textsf{eff}}$ is a constant of integration that is arbitrarily
disposable (as in classical General Relativity), and, hence, is independent of any
fundamental value assigned to the vacuum energy.
This solves the major problem of a huge contradiction between the calculated vacuum energy and the cosmologically observed effective value of the cosmological constant. Any huge $\Lambda_{\rm vac}$ is
powerless to affect cosmology, or indeed the solar system, as the
vacuum energy will not affect spacetime geometry. \\

It has been suggested that there are problems with this proposal, because on the face of it, it also disempowers the scalar field potential from having any effect on inflationary dynamics. This paper has shown that concern is unjustified: that potential is indeed able to influence inflationary dynamics as in the standard theory. It is just its zero point value that does not matter - as is the case in general for potential energy. \\

Is there any observational evidence for the approach based in the TFE discussed above, as opposed to the standard approach based in the EFE? Yes indeed. They give the same results for everything except that, unlike the TFE,  the use of the EFE in conjunction  with the vacuum energy estimates leads to the conclusion that the effective cosmological constant should be 70 orders of magnitude larger than observed.  The data decisively prefer the TFE over the EFE.\\

\textbf{Acknowledgement}: I thank Jean-Philippe Uzan and Roy Maartens for very helpful comments.


\end{document}